\documentclass[10pt]{article}

\usepackage{amsmath}

\usepackage{array}

\usepackage{appendix}

\usepackage{tocloft}                   

\usepackage{graphicx}

\usepackage{amsfonts}

\usepackage{amssymb}

\usepackage{mathrsfs}

\usepackage{yfonts}

\usepackage{euscript}

\usepackage{centernot}                 

\usepackage{ifsym}                     

\usepackage{upgreek}

\usepackage{mathtools}

\usepackage{color}

\usepackage{slantsc}
\usepackage{calligra}

\usepackage{bbold}          

\usepackage[T1]{fontenc}

\usepackage{epsf}

\usepackage{latexsym}

\usepackage{tipa}

\usepackage{makeidx}

\makeindex

\def\be{\begin{equation}}

\def\ee{\end{equation}}

\def\beq{\begin{equation}}

\def\eeq{\end{equation}}

\def\bea{\begin{eqnarray}}

\def\eea{\end{eqnarray}}

\def\!{\hspace{-1.6667em}}

\def\m{\mbox{ }}

\def\mma {\m , \m \m }

\def\!{\hspace{-1.6667em}}

\def\c{\cite}

\def\n{\noindent}

\def\f{\footnote}

\def\u{\underline}

\def\s{\stackrel}

\def\es{\m = \m}

\def\:={\m := \m}

\def\=:{\m =: \m}

\def\biw{\mbox{\boldmath$w$}}

\def\biz{\mbox{\boldmath$z$}}

\def\mA{\mbox{A}}

\def\mI{\mbox{I}}                        

\def\mT{\mbox{T}} 

\def\mU{\mbox{U}}                        

\def\mV{\mbox{V}}

\def\ms{\mbox{s}}

\def\urho{{\underline{\rho}}}

\def\sa{\mbox{\scriptsize a}}

\def\sC{\mbox{\scriptsize C}}

\def\sV{\mbox{\scriptsize V}}

\def\tC{\mbox{\tiny C}}

\def\tS{\mbox{\tiny S}}

\def\bigupalpha{\mbox{\Large$\alpha$}}

\def\cr{\mbox{\scriptsize{\bf $\mbox{ } \times \mbox{ }$}}}

\def\sumi2{\sum\mbox{}_{\mbox{}_{\mbox{\scriptsize $i$=1}}}^2}

\def\sumi3{\sum\mbox{}_{\mbox{}_{\mbox{\scriptsize $i$=1}}}^3}

\def\sumin{\sum\mbox{}_{\mbox{}_{\mbox{\scriptsize $i$=1}}}^{n}}

\def\sumABcycles3{\sum\mbox{}_{\mbox{}_{\mbox{\scriptsize cycles $A,B$=1}}}^{3}}

\def\sumCDcycles3{\sum\mbox{}_{\mbox{}_{\mbox{\scriptsize cycles $C,D$=1}}}^{3}}

\def\sumj3{\sum\mbox{}_{\mbox{}_{\mbox{\scriptsize $j$=1}}}^3}

\def\sumk3{\sum\mbox{}_{\mbox{}_{\mbox{\scriptsize $k$=1}}}^3}

\def\prodiA1{\prod\mbox{}_{\mbox{}_{\mbox{\scriptsize $i$=1}}}^{A - 1}}

\def\bigtimes{\mbox{\Large $\times$}}

\def\d{\textrm{d}}                                                  

\def\sFrP{\mbox{\scriptsize $\mathfrak{P}$}}
												  
\def\sFrR{\mbox{\scriptsize $\mathfrak{R}$}}                   
\def\FrS{\mbox{\Large $\mathfrak{s}$}}                         

\def\Hilb{\mbox{{\boldmath$\mathfrak{H}$}ilb}}                 
\def\FrQ{\mbox{\Large $\mathfrak{q}$}}                               
\def\sFrQ{\mbox{\large $\mathfrak{q}$}}                              
\def\Phase{\mbox{{\boldmath$\mathfrak{P}$}hase}}                     

\def\bFrR{\mbox{\boldmath$\mathfrak{R}$}}                            
\def\Rig-Phase{\bFrR\mbox{ig-}\Phase}                                
                                                                													   
\def\nFrr{\mbox{\normalsize $\mathfrak{r}$}}                              

\def\FrP{\mbox{\Large $\mathfrak{p}$}}                                 
\def\sFrP{\mbox{\large $\mathfrak{p}$}}                                

\def\FrR{\mbox{\boldmath$\mathfrak{R}$}}                             

\def\sFrR{\mbox{\scriptsize\boldmath$\mathfrak{R}$}}                 
\def\bFrM{\mbox{\boldmath${\mathfrak{M}}$}}                             

\def\Positive-Modespace{\mbox{{\boldmath$\mathfrak{M}$}odespace$^+$}}

\def\POSITIVE-MODESPACE{\mbox{{\boldmath$\mathfrak{M}$}ODESPACE$^+$}}

\def\Kin-Hilb{\mbox{{\boldmath$\mathfrak{K}$}in-\Hilb}}                     

\def\Mid-Hilb{\mbox{{\boldmath$\mathfrak{M}$}id-\Hilb}}                     

\def\Dyn-Hilb{\mbox{{\boldmath$\mathfrak{D}$}yn-\Hilb}}                     

\def\5Star{\mbox{\Large$\star$}}              

\begin{document}

\begin{titlepage}

\begin{center}

\vspace{0.1in}

\Large{\bf Quadrilaterals in Shape Theory. II.} \normalsize

\vspace{0.1in}

\large{\bf Alternative Derivations of Shape Space: Successes and Limitations} \normalsize 

\vspace{0.1in}

{\large \bf Edward Anderson$^*$}
-
\vspace{.2in}

\end{center}

\begin{abstract}

We show that the recent derivation that triangleland's topology and geometry is $\mathbb{S}^2$ from Heron's formula does not extend to quadrilaterals 
by considering Brahmagupta, Bretschneider and Coolidge's area formulae. 
That $N$-a-gonland is more generally $\mathbb{CP}^{N - 2}$ (with $\mathbb{CP}^1 = \mathbb{S}^2$ recovering the triangleland sphere)  
follows from Kendall's extremization that is habitually used in Shape Theory, or the generalized Hopf map. 
We further explain our observation of non-extension in terms of total area not providing a shape quantity for quadrilaterals. 
It is rather the square root of of sums of squares of subsystem areas that provides a shape quantity; we clarify this further in representation-theoretic terms.   
The triangleland $\mathbb{S}^2$ moreover also generalizes to $d$-simplexlands being $\mathbb{S}^{d(d + 1)/2 - 1}$ topologically by Casson's observation.  
For the 3-simplex - alias tetrahaedron - while volume provides a shape quantity and is specified by the della Francesca--Tartaglia formula, 
the analogue of finding Heron eigenvectors is undefined. 
$d$-volume moreover provides a shape quantity for the $d$-simplex, specified by the Cayley--Menger formula generalization of the Heron and della Francesca--Tartaglia formulae.
While eigenvectors can be defined for the even-$d$ Cayley--Menger formulae, 
the dimension count does not however work out for these to provide on-sphere conditions.  
We finally point out the multiple dimensional coincidences behind the derivation of the space of triangles from Heron's formula. 
This article is a useful check on how far the least technically involved derivation of the smallest nontrivial shape space can be taken. 
This is significant since Shape Theory is a futuristic branch of mathematics, with substantial applications in both Statistics (Shape Statistics) 
and Theoretical Physics (Background Independence: of major relevance to Classical and Quantum Gravitational Theory).

\m 

\end{abstract}

\n PACS: 04.20.Cv, 02.40.-k, Physics keywords: Background Independence, configuration spaces, 4-body problem.  

\m

\n Mathematics keywords: Euclidean, Projective and Differential Geometry, Shape Theory, spaces of simplexes, Distance Geometry, Shape Statistics. 

\vspace{0.1in}
  
\n $^*$ Dr.E.Anderson.Maths.Physics *at* protonmail.com

\section{Introduction}\label{Intro-III}

\n The {\it carrier space} $\bFrM^d$, 
                   alias {\it absolute space} in the physical context is an at least provisional model for the structure of space.
\n{\it Constellation space} (reviewed in \cite{FileR}) is then the product of $N$ copies of this carrier space, 
\be 
\FrQ(\bFrM^d, \, N) \es \bigtimes_{I = 1}^N \bFrM^N
\label{Stell}
\ee 
modelling $N$ points on $\bFrM^d$, or, if these points are materially realized, $N$ particles (classical, nonrelativistic). 
In {\it Kendall's Shape Theory} \cite{Kendall84, Kendall89, Kendall} 
(see also \cite{Small, JM00, FORD, GT09, FileR, Bhatta, MIT, DM16, PE16, KKH16, I, II, III, Shape-Theory, IV}), 
we consider this for $\bFrM^d = \mathbb{R}^d$ furthermore with similarities $Sim(d)$ quotiented out. 
The corresponding {\it shape spaces} -- reduced configuration spaces -- are thus of the form  
\be 
\FrS(d, N) \es \frac{\times_{I = 1}^N \mathbb{R}^d}{Sim(d)} \m .
\ee 
Kendall's work remains rather more familiar in the Shape Statistics literature \cite{Kendall84, Kendall89, Small, Kendall, JM00, Bhatta, DM16, PE16}, 
though related work in other fields has also appeared in e.g.\  Mechanics and Molecular Physics \cite{LR95, LR97, Montgomery, FORD, FileR, Shape-Theory}, 
and in \cite{FORD, FileR, APoT, ABook} as regards modelling some aspects of General Relativity's Background Independence 
\cite{Battelle, DeWitt, Kuchar92, I93, APoT3, ABook, PE-1, DO-2}.  

\m 

\n As we outline in Sec 2, diagonalizing the usual Heron's formula was recently demonstrated \cite{A-Heron} to suffice as first principles to derive both 

\m 

\n 1) `{\it Kendall's Little Theorem}' that the shape space of triangles \cite{Kendall84, Kendall89, Kendall} is 
\be 
\FrS(2, 3) = \mathbb{S}^2 
\ee 
\n 2) the Hopf map \cite{Hopf}, in the extended ${\cal H}_{\sC}$ sense of Fig 1.a-b).  

\m

{            \begin{figure}[!ht]
\centering
\includegraphics[width=1.0\textwidth]{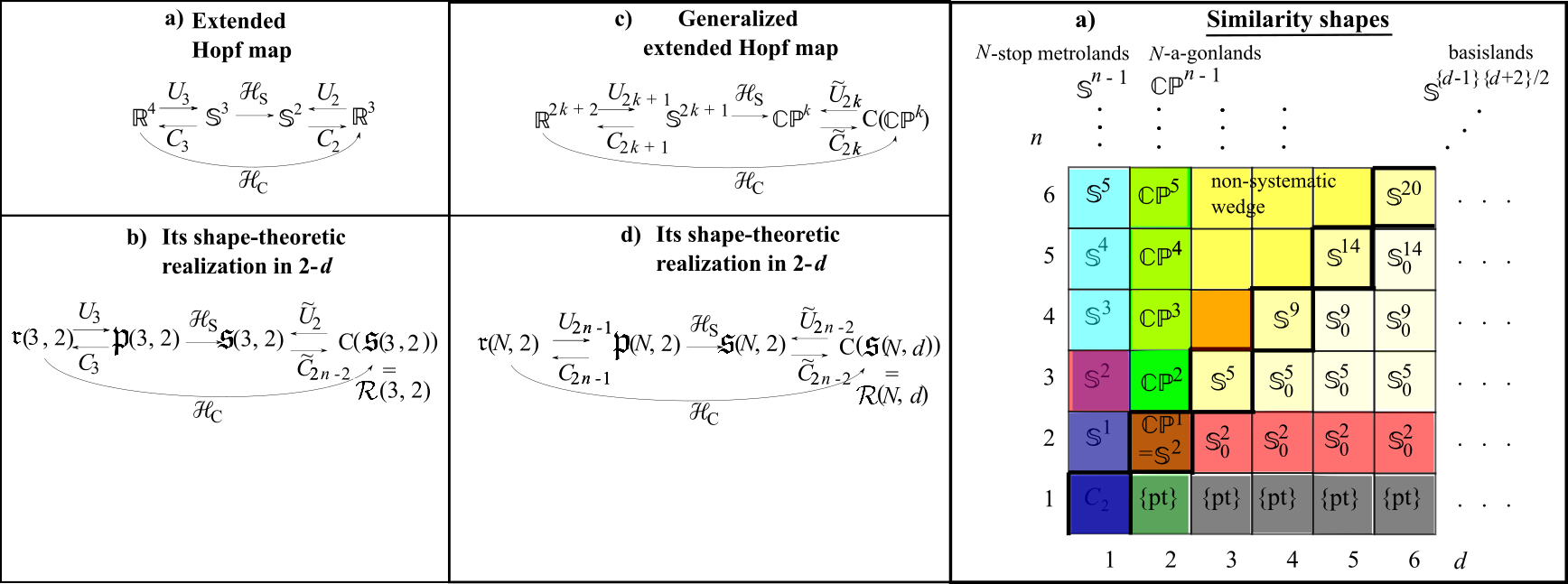}
\caption[Text der im Bilderverzeichnis auftaucht]{        \footnotesize{
a)-d) Hopf map ${\cal H}_{\tS}$ from $\mathbb{S}^{2 \, k + 1}$ to $\mathbb{CP}^k$, 
including mapping the natural ambient $\mathbb{R}^{2 \, k + 2}$ for the $\mathbb{S}^{2 \, k + 1}$ to the cone over  $\mathbb{CP}^k$
 by the Hopf map ${\cal H}_{\tC}$, where the $\sC$ now stands for `cone' rather than `Cartesian'.
$\widetilde{C}_k$ map an arbitrary manifold to its topological and geometrical cone, 
whereas $\widetilde{U}_k$ are unit-scaling maps from the cone back to the manifold being coned over.
$\nFrr(d, N) := \sFrQ(d, N)/Tr(d)$ is {\it relative space}, 
$\sFrP(d, N) := \sFrQ(d, N)/Tr(d) \rtimes Dil$ is Kendall's \cite{Kendall84, Kendall} {\it preshape space} and 
$\sFrR(d, N) :=  \sFrQ(d, N)/Eucl(d)$ is {\it relational space}.   
e) The $(d, N)$ grid of shape spaces at the topological level, with 1-$d$, 2-$d$ and diagonal Casson series marked.
$\mathbb{S}_0^k$ denotes the $k$-dimensional hemisphere.}  } 
\label{Hopf-Map-Casson} \end{figure}           }

\n The derivation in question, as recapped in Sec 2, is via Linear Algebra reformulation in terms of the Heron--Buchholz matrix \cite{Buchholz}, 
its diagonalization \cite{A-Heron} the total moment of inertia $\iota$ being among the eigenvectors, and division by it resulting in the on-$\mathbb{S}^2$ condition. 

\end{titlepage}

\n The current Article considers the extent to which these results generalize to Shape Theory with other spatial dimensions $d$ and point-or-particle numbers $N$.    
There are two directions of generalization (Fig 1.e) away from triangleland in 2-$d$

\m 

\n 1) $N$-a-gons, for which {\it Kendall's Theorem} (\cite{Kendall84, Kendall} and the Appendix) gives that 
\be
\FrS(N, 2) = \mathbb{CP}^{N - 2}
\ee 
both topologically and metrically, with standard Fubini--Study metric.  
This generalization makes sense via the accidental topological-and-geometrical relation 
\be 
\mathbb{CP}^1 = \mathbb{S}^2                    \m .  
\ee 
\n 2) $d$-simplexes, for which Casson \cite{Kendall} showed that, at the topological level, 
\be 
\FrS(d, d + 1) = \mathbb{S}^{d(d + 1)/2 - 1}    \m .  
\ee 
\n For the quadrilaterals, we show in Sec 3 that Coolidge's area formula is sequentially more satisfactory than 
                                                 Brettschneider's
		                                     and Brahmagupta's for this application, but still falls short of requirements to provide a derivation of the topology and geometry 
											 of the space of quadrilaterals.  
What works instead for $N$-a-gons is Kendall's extremization or the generalized Hopf map in the extended sense ${\cal H}_{\sC}$ of Fig 1.c)-d) (both outlined in the Appendix), 
or a Mechanics reduction for which references are given in the Conclusion. 

\m 

\n We moreover further explain our observation of non-extension in Sec 4, in terms of total mass-weighted area $\alpha$ per unit moment of inertia $\iota$, 
\be 
\frac{  \bigupalpha  }{  \iota  } 
\ee  
not being a shape quantity \cite{FileR, Quad-I} for quadrilaterals. 
It is rather 
\be
\frac{    \sqrt{  \sum_{A = 1}^3\bigupalpha_A^2  }    }{    \iota   }     
\ee  
\cite{Quad-I} which is a shape quantity, the sum being over the three two-Jacobi-vector subsystems supported by the quadrilateral. 
We further clarify the reason for this object's featuring in the theory of quadrilaterals in Representation-Theoretic terms.   

\m 

\n We next consider whether there is a Casson's Theorem generalization of the Heron proof of `Kendall's Little Theorem' in Sec 5.  
For the 3-simplexes alias tetrahaedrons (for which Sec 8 provides the notation), volume both has a general formula in terms of separations 
-- the della Francesca--Tartaglia formula \cite{DF, Tartaglia} -- and provides a shape quantity 
\be 
\frac{\cal V}{\iota^{3/2}}
\ee 
(for ${\cal V}$ the mass-weighted volume).   
We provide the analogue of the Heron--Buchholz \cite{Buchholz} matrix reformulation in terms of a `Tartaglia 3-array'. 
We then observe that eigenvectors are however not defined for this, by which the Heron derivation of the triangleland shape space does not generalize to tetrahaedronland either.  

\m 

\n In Sec 6, we finally consider the general $d$-simplexes, 
for which the $d$-volume is given by the Cayley--Menger generalization \cite{Cayley, Menger, Blumenthal53, CH88} of the della Francesca--Tartaglia formula. 
$d$-volume moreover provides a shape quantity 
\be 
\frac{{\cal V}_d}{\iota^{d/2}}                         \m , 
\ee 
and eigenvectors can be defined for the even-$d$ cases. 
The dimension count does not however work out for this to give on-sphere conditions corresponding to Casson's result for the topology of the spaces of $d$-simplexes. 
These observations made, we point to the string of dimensional coincidences behind the Heron derivation in Secs 4 and 6.

\section{Ouline of Triangleland shape space from Heron's formula}
%
{            \begin{figure}[!ht]
\centering
\includegraphics[width=0.27\textwidth]{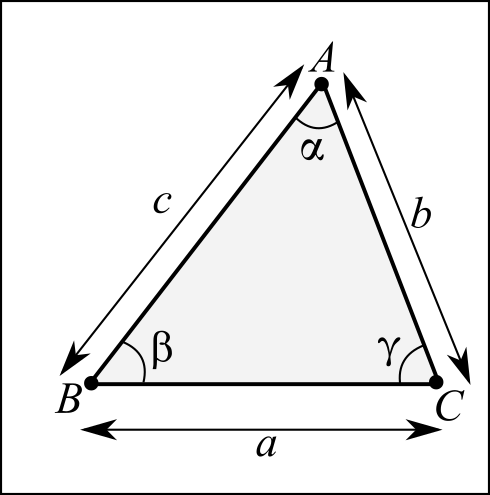}
\caption[Text der im Bilderverzeichnis auftaucht]{        \footnotesize{a) Labelling of vertices and edges of the triangle.  }  } 
\label{Triangle-Notation} \end{figure}           }

\n{\bf Definition 1} Consider an arbitrary triangle $\triangle\,ABC$, denoted as in Fig \ref{Triangle-Notation}.   
Using $s_i$, $i = 1$ to $3$ to denote $a$, $b$, $c$ will also be useful for us.   

\m 

\n{\bf Definition 2} The {\it semi-perimeter} is  
\be
\ms  \m := \m  \frac{a + b + c}{2} 
     \m  = \m  \frac{1}{2}\sum_{i = 1}^{3} s_i                       \m .  
\ee
\n{\bf Definition 3} We use $\mA(ABC) = \mA$ for short when unambiguous, to denote the area of $\triangle\,ABC$.

\m 

\n{\bf Proposition 1 (Heron's formula)}  
\be
\mA  \es \sqrt{s(s - a)(s - b)(s - c)}                               \m .  
\ee
\n{\bf Remark 1} Squaring, introducing the tetra-area variable 
\be 
\mT  \:=  4 \, \mA                                                   \m ,  
\ee 
and expanding out, 
\be   
\mT^2 \es (2 \, a^2 b^2 - c^4) \mbox{ + cycles}                       \m . 
\label{Exp-Heron} 
\ee
\n{\bf Remark 2} This can furthermore be recast in Linear Algebra terms as the quadratic form,   
\be 
\mT^2 = H_{ij} {\sigma^i} {\sigma^j}
\label{Heron-form}
\ee 
for {\it squared separations} (for triangles, separations are just side-lengths) 3-vector 
\be 
\sigma^i := s^{i \, 2} \m \mbox{ (no sum) }
\ee 
and  `{\it Heron--Buchholz matrix}' \cite{Buchholz}
\be
\underline{\underline{H}} := 
\mbox{$\frac{1}{3}$}   
\mbox{\Huge(} \s{  \s{  \mbox{\scriptsize --1}}{\mbox{\scriptsize 1}}}    {\mbox{\scriptsize 1}}                            \m 
              \s{  \s{  \mbox{\scriptsize $1$}}  {\mbox{\scriptsize --1}}}{\mbox{\scriptsize 1}}                            \m 
              \s{  \s{  \mbox{\scriptsize $1$}}  {\mbox{\scriptsize 1}}}    {\mbox{\scriptsize --1}}       \mbox{\Huge)}    \m .
			  \label{H-B}
\ee
{\bf Remark 3} Setting  
\be
0  =  \mbox{det}\left(\underline{\underline{H}} - \lambda \, \underline{\underline{I}} \right)                              \m ,
\ee
the eigenvalues are $\lambda = 1$ with multiplicity 1, and $\lambda = -2$ with multiplicity 2.

\m

\n The corresponding orthonormal eigenvectors are, respectively, 
\be 
\mbox{$\frac{1}{\sqrt{3}}$}    \mbox{\Huge(}    \s{ \mbox{\scriptsize 1}  } {  \s{  \mbox{\scriptsize 1}     } { \mbox{\scriptsize 1}      }  }  \mbox{\Huge)}        \mma 
\mbox{$\frac{1}{\sqrt{2}}$}    \mbox{\Huge(}    \s{ \mbox{\scriptsize 1}  } {  \s{  \mbox{\scriptsize --1}  } { \mbox{\scriptsize 0}      }  }       \mbox{\Huge)}    \mma
\mbox{$\frac{1}{\sqrt{6}}$}    \mbox{\Huge(}    \s{ \mbox{\scriptsize 1}  } {  \s{  \mbox{\scriptsize 1}     } { \mbox{\scriptsize --2}  }  }   \mbox{\Huge)}         \m .  
\label{H-evectors}
\ee 
\n{\bf Remark 4} The diagonalizing variables are thus 
\be
\overline{a}^2  \es  \frac{a^2 + b^2 + c^2}{\sqrt{3}}       \mma 
\overline{b}^2  \es  \frac{a^2 - b^2}{\sqrt{2}}             \m , \m \mbox{ and }
\overline{c}^2  \es  \frac{a^2 + b^2 - 2 \, c^2}{\sqrt{2}}  \m .   
\ee 
In terms of these, Heron's formula also takes the following form. 

\m

\n{\bf Diagonal Heron formula} 
\be 
\mT  \es \sqrt{  \overline{a}\mbox{}^4 - 2(\overline{b}\mbox{}^4 + \overline{c}\mbox{}^4)  }      \m .  
\label{T}
\ee 
Then upon introducing the rescaled {\it ratio variables} \cite{III}
\be 
X \m := \m \sqrt{2}  \left(  \frac{\overline{b}}{\overline{a}}  \right)^2 
  \m  = \m \frac{\sqrt{3}(a^2 - b^2)  }{  a^2 + b^2 + c^2 }                                     \mma 
Y \m := \m \frac{\mT}{\overline{a}^2}                                                           \mma 
Z \m := \m \sqrt{2}  \left(  \frac{  \overline{c}  }{  \overline{a}  }  \right)^2 
  \m  = \m \frac{  a^2 + b^2 - 2 \, c^2  }{  a^2 + b^2 + c^2 }                                  \mma   
\label{RR}
\ee 
whose denominator is proportional to the moment of inertia, (\ref{T}) becomes 
\be 
X^2 + Y^2 + Z^2 = 1                                                                             \m ,  
\ee 
the on 2-sphere condition.  

\m

\n{\bf Corollary 1} This gives `Kendall's Little Theorem': that the space of triangles is a 2-sphere. 

\m

\n{\bf Corollary 2} We furthermore identify $X$, $Y$, $Z$ as Hopf quantities; in terms of normalized \cite{I} mass-weighted relative Jacobi vectors \cite{Marchal}, 
these take the familiar 
\be 
X = 2 \, \u{\nu}_1 \cdot \u{\nu}_2   \m , 
\label{X}
\ee  
\be 
Y = 2 \, \u{\nu}_1 \cr \u{\nu}_2     \m , 
\label{Y}
\ee 
\be 
Z =    {\nu_2}^2 - {\nu_1}^2         \m .  
\label{Z}
\ee 
In this way, Heron's formula provides the Hopf map as well as `Kendall's Little Theorem', by which the above derivation gains its Heron--Hopf--Kendall monicker. 
That squares of expressions (\ref{X}-\ref{Z}) add to 1 is standard: 
$$ 
X^2 + Y^2 + Z^2  \es  4  (\nu_{1x} \nu_{2x} + \nu_{1y} \nu_{2y} )^2 + 4 (\nu_{1x} \nu_{2y} - \nu_{1y} \nu_{2x} )^2 + ({\nu_{2x}}^2 + {\nu_{2y}}^2 - {\nu_{1x}}^2 - {\nu_{1y}}^2 )^2
$$
$$ 
\es 4 {\nu_{1x}}^2 {\nu_{2x}}^2 + 4 {\nu_{1y}^2 \nu_{2y}^2} + 8 {\nu_{1x}} {\nu_{2x}} {\nu_{1y}} {\nu_{2y}} +
    4 {\nu_{1x}}^2 {\nu_{2y}}^2 + 4 {\nu_{1y}^2 \nu_{2x}^2} - 8 {\nu_{1x}} {\nu_{2x}} {\nu_{1y}} {\nu_{2y}} + 
$$
$$
	{\nu_{2x}}^4 + {\nu_{2y}}^4 + {\nu_{1x}}^4 + {\nu_{1y}}^4 +
    2 ( {\nu_{2x}}^2 {\nu_{2y}}^2 - {\nu_{1x}}^2 {\nu_{2x}}^2 - {\nu_{2x}}^2 {\nu_{1y}}^2 - {\nu_{2y}}^2 {\nu_{1x}}^2 - {\nu_{2y}}^2 {\nu_{1y}}^2 + {\nu_{1x}}^2 {\nu_{1y}}^2)
$$ 
\be  
\es {\nu_{2x}}^4 + {\nu_{2y}}^4 + {\nu_{1x}}^4 + {\nu_{1y}}^4 + 
    2 ( {\nu_{2x}}^2 {\nu_{2y}}^2 + {\nu_{1x}}^2 {\nu_{2x}}^2 + {\nu_{2x}}^2 {\nu_{1y}}^2 + {\nu_{2y}}^2 {\nu_{1x}}^2 + {\nu_{2y}}^2 {\nu_{1y}}^2 + {\nu_{1x}}^2 {\nu_{1y}}^2) 
\es ({\nu_1}^2 + {\nu_2}^2)^2 
\es 1	                             \m .  
\ee    
\n{\bf Remark 9} Hopf quantities are the form taken by triangleland's shape quantities \cite{+Tri, FileR, III}. 
$X$ is geometrically an {\it anisoscelesness}: a measure of departure from isoscelesness, the mass-weighted area per unit moment of inertia. 
$Z$ is geometrically an {\it ellipticity}: a quantifier of whether the triangle is tall or flat.  
Anisoscelesness and ellipticity can moreover now be interpreted as the two eigenvectors of the Heron map $\underline{\underline{H}}$ other than the total moment of inertia. 
This comment is useful as regards the current Article's assessment of whether quadrilateral area formulae similarly involve quadrilateralland's shape quantities \cite{IV}.   


\section{Quadrilateral area formulae}
%
{            \begin{figure}[!ht]
\centering
\includegraphics[width=0.45\textwidth]{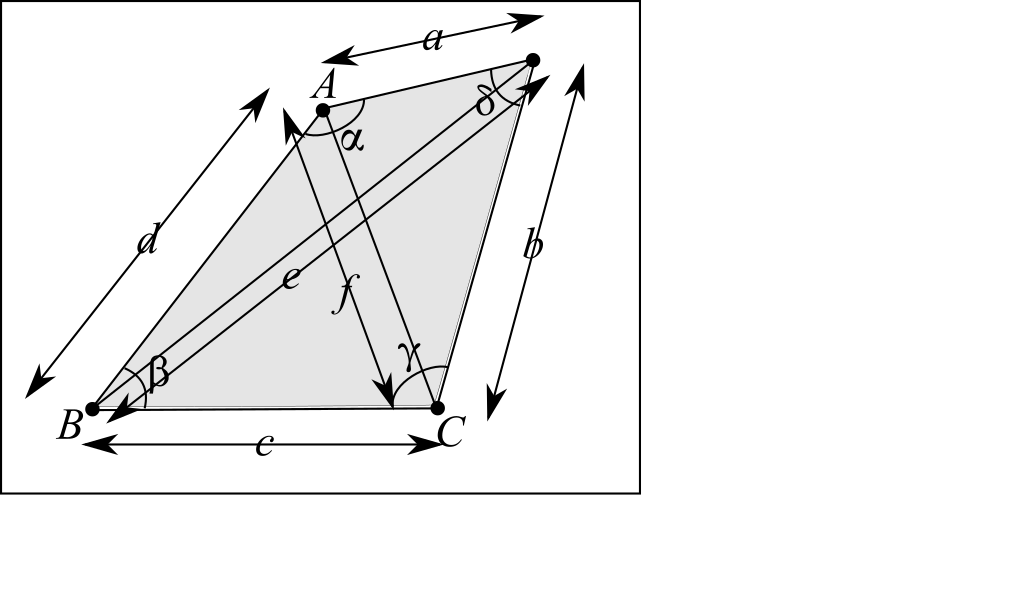}
\caption[Text der im Bilderverzeichnis auftaucht]{        \footnotesize{Labelling of vertices, edges, angles and diagonals of the quadrilateral. }  } 
\label{Quadrilateral-Notation} \end{figure}           }

\n{\bf Definition 1} Consider an arbitrary quadrilateral denoted as in Fig \ref{Quadrilateral-Notation}.   

\m 

\n We consider in particular formulae for the area $\mA$ of quadrilaterals. 

\m 

\n{\bf Proposition 1 (Brahmagupta's formula)} (7th century A.D. India \cite{Brahmagupta, CG67}). For a cyclic quadrilateral, 
\be 
\mA  =  \sqrt{(s - a)(s - b)(s - c)(s - d)}                                       \m . 
\label{Brahma-1}
\ee 
\n{\bf Remark 1} Setting $d = 0$, this returns Heron's formula.

\m 

\n{\bf Remark 2} In expanded form, 
\be 
\mT^2  \es  4 \, ( a \, b + c \, d )^2 - (a^2 - c^2 + b^2 - d^2)^2  
       \es  {S_2}^2 - 2 \, S_4 + 8 \, S_{\Pi}                                         \m .
\label{Brahma-2}
\ee 
Here 
\be 
S_{\sa}  \es  \sum_{i = 1}^4 {s_i}^{\sa}                                          \m  
\ee 
and 
\be 
S_{\Pi} \es \prod_{i = 1}^4 {s_i}                                                 \m . 
\ee 
Unlike for Heron's formula, the second expanded form precludes dependence on ${s_i}^2$ alone. 
This feature simplified repackaging Heron's formula as a quadratic form in squared variables \cite{A-Heron}.  

\mbox{ }

\n{\bf Remark 3} The formula being solely for cyclic quadrilaterals is also a limitation.  

\mbox{ }

\n{\bf Remark 4} For a general quadrilateral, SSSS -- four sides -- is not sufficient data \cite{IV}. 
Thus no area formula exists without bringing in some further datum. 

\m 

\n{\bf Remark 5} One alternative in bringing in more data is to involving angles between sides as data.  
This leads to the following. 

\m

\n{\bf Proposition 2 (Bretschneider's formula)} \cite{Bretschneider} (1842) For a quadrilatral, 
\be 
\mA  \es  \sqrt{(s - a)(s - b)(s - c)(s - d) - a \, b \, c \, d \, \mbox{cos}^2\left(\frac{\alpha + \gamma}{2}\right)}  \m . 
\label{Bret-1}
\ee
\n{\bf Remark 6} As indicated in Fig 3, $\alpha$ and $\gamma$ are opposite angles. 
Thus in the cyclic case, their sum is $\pi$ by an elementary Theorem of Euclid's. 
\be
\mbox{cos} \, \frac{\pi}{2} = 0 
\ee 
then accounts for the last term of this vanishing in Brahmagupta's formula.  

\m 

\n{\bf Remark 7} Thus we interpret the new term conceptually as 
\be 
a \, b \, c \, d \, \mbox{cos}^2\left(\frac{\alpha + \gamma}{2}\right) =: \mbox{cyclator : acyclicity contribution to area}    \m . 
\label{Acyc-1}
\ee 
\n{\bf Remark 8} There are however two reasons why Bretschneider's area formula does not provide a useful extension of Heron's formula for shape-theoretic use. 

\m 

\n{\bf Reason 1} Involvement of the conceptually heterogeneous angular information, moreover in multiplicative fashion on one factor. 

\m 

\n{\bf Reason 2} $a \, b \, c \, d$ does not depend on the sides via their squares.  

\m 

\n These matters are moreover remedied by making use instead of {\sl diagonal length data}.   
This is homogeneous with side length data through both constituting separation data.
Indeed, in separational alias Lagrangian and dual constellational approaches \c{I} -- an accurate description of Shape Theory -- 
no distinction is to be made between sides and diagonals, so modelling purely in terms of unqualified separations remains faithful to this. 

\m 

\n{\bf Proposition 3 (Coolidge's formula)} \cite{Coolidge} (1939) For a quadrilateral,
\be 
\mA  \es \sqrt{  (s - a)(s - b)(s - c)(s - d) - \frac{1}{4}(a \, c + b \, d + e \, f)(a \, c + b \, d - e \, f)  }  \m . 
\label{Cool-1}
\ee
{\bf Remark 9} This reduces to Brahmagupta's formula (\ref{Brahma-1}) iff the quadrilateral is cyclic.
This is by the last term's last factor embodying Ptolemy's Theorem \cite{Ptolemy}, \cite{CG67}
\be 
e \, f = a \, c = b \, d \m \Leftrightarrow \m  
ABCD  \m \mbox{ is cyclic}                                                                                        \m , 
\ee
and the last term's first factor offering not other ways of being zero. 
We have thus found another formulation for the cyclator, 
\be 
\mbox{cyclator}  \es  \frac{1}{4}(a \, c + b \, d + e \, f)(a \, c + b \, d - e \, f)    \m . 
\ee 
\n{\bf Corollary 1} The {\it expanded version of Coolidge's formula} is moreover  
\be 
\mT^2 = e^2 f^2 - (a^2 - b^2 + c^2 - d^2)^2                                                                         \m , 
\label{Cool-2}
\ee 
which is additionally manifestly of the form 
\be 
\mA = f(\mbox{separations}^2 \m \mbox{alone})                                                                       \m . 
\ee 
\n{\bf Remark 10} Coolidge's formula thus generalizes Heron's formula in ways 1) and 2). 
Thus it admits the following re-expression. 

\m 

\n{\bf Corollary 2} The tetra-area squared is given by the {\it Coolidge quadratic form} 
\be 
\mT^2 = C_{ij} \sigma^i \sigma^j                                                                                             \m .  
\ee
Here, $\sigma^i$ is the $\mbox{separation}^2$ 6-vector 
\be
\sigma^i = (a^2, b^2, c^2, d^2, e^2, f^2)                                                                             \m ,
\ee 
on the `separation space' of Distance Geometry \c{Blumenthal53, CH88}.
Also Fig \ref{Coolidge}'s $C_{ij}$ is the $6 \times 6$ {\it Coolidge matrix}. 
%
{            \begin{figure}[!ht]
\centering
\includegraphics[width=0.25\textwidth]{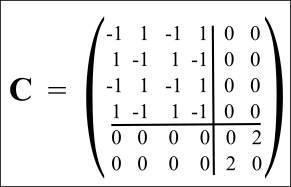}
\caption[Text der im Bilderverzeichnis auftaucht]{        \footnotesize{The Coolidge matrix}  } 
\label{Coolidge} \end{figure}           }

\m

\n{\bf Remark 11} The Coolidge matrix's eigenvalues and eigenvectors -- solving 
\be 
(  C_{ij} - \lambda \, \delta_{ij}  ) x_j = 0 
\ee 
-- are as follows. 
We can solve this separately for its $2 \times 2$ and $4 \times 4$ block, obtaining the eigenspectrum 
\be 
\lambda = 0 \m \mbox{ (multiplicity 3)}, -2, 2, 4   \m . 
\ee 
The corresponding eigenvectors are 
\be 
\mbox{$\frac{1}{\sqrt{2}}$}( 0, \, 0, \,  0, \, 0, \, 1, \, 1)            \mma 
\mbox{$\frac{1}{\sqrt{2}}$}( 0, \, 0, \, 0, \, 0, \, 1, \, -1)            \mma
\mbox{$\frac{1}{2}$}       (-1, \, 1, \, -1,  \, 1, \, 0,  \, 0)            \mma
\ee 
for -2, 2 and 4 respectively and e.g.\ 
\be 
\mbox{$\frac{1}{\sqrt{2}}$} (1, \, 1, \, 1,  \, 1, \, 0, \, 0)            \mma
\mbox{$\frac{1}{\sqrt{2}}$} (1, \, 0, \, -1, \, 0, \, 0, \,  0)            \mma
\mbox{$\frac{1}{2}$}        (0, \, 1,  \, 0, \, -1, \, 0,  \, 0)            \mma
\ee 
for the 3-$d$ 0 eigenspaces. 

\m 

\n{\bf Remark 12} These eigenvectors admit the following geometrical interpretation.  
\be 
\mbox{$\frac{1}{\sqrt{2}}$}(e^2 + f^2)        \es  \mbox{$\frac{1}{\sqrt{2}}$}(\mbox{diagonal squares sum})  
                                     \es  \mbox{$\frac{1}{\sqrt{2}}$}(\mbox{diagonal size})                                          \m . 
\ee 
\be 
\mbox{$\frac{1}{\sqrt{2}}$}(e^2 - f^2)        \es  \mbox{$\frac{1}{\sqrt{2}}$}(\mbox{diagonal squares difference})  
                                     \es  \mbox{$\frac{1}{\sqrt{2}}$}(\mbox{diagonal ellipticity})                                   \m . 
\ee
\be 
\mbox{$\frac{1}{2}$}(-a^2 + b^2 - c^2 + d^2)  \es  \mbox{$\frac{1}{2}$}(\mbox{alternating side squares sum})  
                                     \es  \mbox{$\frac{1}{\sqrt{2}}$}(\mbox{difference of adjacent sides' ellipticities})            \m . 
\ee
\be 
\mbox{$\frac{1}{2}$}(a^2 + b^2 + c^2 +  d^2)  \es  \mbox{$\frac{1}{2}$}(\mbox{unit-mass moment of inertia})                         \m . 
\ee
\be 
\mbox{$\frac{1}{\sqrt{2}}$}(a^2 - c^2)        \es  \mbox{$\frac{1}{\sqrt{2}}$}(\mbox{ellipticity of an opposite-sides pair})         \m . 
\ee
\be 
\mbox{$\frac{1}{\sqrt{2}}$}(b^2 - d^2)        \es  \mbox{$\frac{1}{\sqrt{2}}$}(\mbox{ellipticity of the other opposite-sides pair})  \m . 
\ee
Note the dimensional mismatch: at best this gives a 5-sphere, without recourse for a second restriction down to $\mathbb{CP}^2$.

\section{Shape quantities for the quadrilateral}

Let us first introduce Kendall's preshape space \cite{Kendall84, Kendall}: the result of quotienting constellation space (\ref{Stell}) by the dilatations: 
the translations and dilations in the semidirect product \cite{Cohn} form 
\be 
Dilatat(d) \es  Tr(d) \rtimes Dil \es \mathbb{R}^d \rtimes \mathbb{R}_+  \m .   
\ee 
{\it Preshape space} itself is then 
\be 
\FrP(d, \, N)  \es  \frac{\FrQ(d, \, N)}{Dilatat(d)} 
               \es  \frac{\mathbb{R}^{d, \, N}}{\mathbb{R}^d \rtimes \mathbb{R}_+}  
               \es  \frac{\mathbb{R}^{n \, d}}{\mathbb{R}_+}  
			   \es  \mathbb{S}^{n \, d - 1} 			   \m :
\label{Preshape}			   
\ee 
both topologically and metrically a sphere.  
In 1-$d$, moreover, since there are no continuous rotations, preshape space is equivalent to shape space: 
\be 
\FrS(1, \,  N) = \FrP(1, \, N) = \mathbb{S}^{n - 1}  \m .  
\ee 
\n In 1-$d$, the $n$ relative Jacobi scalars $\rho^i$ are Euclidean invariants. 
Let us normalize these using 
\be 
\rho := \sqrt{\iota} \m ,
\ee 
which furthermore plays the role of \cite{I} radius of preshape space radius 
This leaves us with $n$ quantities\f{This can also be found in the theory of internal rotations, alias democracy transformations in the Molecular Physics literature.} 
\be 
\nu^i  \:=  \frac{\rho^i}{\rho} 
\ee 
subject to the on-$\mathbb{S}^{n - 1}$ condition 
\be 
\sumin{\nu^i}^2 \es 1                                                                                      \m . 
\ee 
This working moreover generalizes for $d$-dimensional space at the level of preshape space \cite{Kendall84, I}, to 
\be 
\u{\nu}^i  \:=  \frac{\u{\rho}^i}{\rho} 
\ee 
being subject to the on-$\mathbb{S}^{n \, d - 1}$ condition 
\be 
\sum_{\Gamma = 1}^{n \, d} || \u{\nu}^i ||^2  \es  1                                                       \m . 
\ee
\n For $d \geq 2$, $\u{\rho}^i\cdot \u{\rho}^j$ are Euclidean invariants. 
There are 
\be 
\frac{n(n + 1)}{2}
\ee 
distinct such, and we can form a matrix out of them: the {\it Euclidean matrix}
\be 
E^{ij}  \:=  (\u{\rho}^i \cdot \u{\rho}^j)                                                                   \m .
\ee 
For $N = 3$, this matrix has 3 independent elements and two invariants,  
\be 
\mbox{Tr}(E)  =  \rho^2 
              =  \iota 
\ee 
and 
\be 
\mbox{Det}(E)    \es    {\rho_1}^2{\rho_2}^2 - (\urho_1 \cdot \urho_2)^2   
                 \es    |\rho_1 \cr \rho_2|^2 
			   \propto  \alpha^2                                                                           \m . 
\ee
The second equality here is Lagrange's identity, and $\alpha$ is the mass-weighted area of the triangle.
For $N = 4$, $E_{ij}$ has six distinct elements and three invariants:  
\be 
\mbox{Tr}(E)   =   \rho^2 
               =   \iota                                                                                     \m , 
\ee 
\be 
\mbox{II}(E)       \es       |\rho_1 \cr \rho_2|^2  \m + \m  \mbox{cycles} 
              \m \propto \m      {\alpha_{12}}^2    \m + \m  \mbox{cycles}                                       \m , 
\ee 
\be 
\mbox{Det}(E)      \es       (\rho_1, \rho_2, \rho_3)^2 
              \m \propto \m  ({\cal V})^2                                                                        \m .
\ee 
The first equality in the last equation is a generalization of Lagrange's identity; this third invariant is moreover zero in 2-$d$.  

\m 

\n Normalizing by $\iota$ gives the {\it similarity matrix} 
\be 
S^{ij}  :=  (\u{\nu}^i \cdot \u{\nu}^j)                                                                           \m . 
\ee
The trace is just a number $N$.  
For $N = 3$, the similarity matrix has just one other invariant, 
\be 
\mbox{Det}(S)      \es           |\nu_1 \cr \nu_2|^2  
               \m \propto \m  \frac{\alpha^2}{\iota^2}                                                           \m . 
\ee 
For $N = 3$, the similarity matrix has two other invariants,  
\be 
\mbox{II}(S)       \es      |\nu_1 \cr \nu_2|^2              \m + \m  \mbox{cycles} 
              \m \propto \m \frac{{\alpha_{12}}^2}{\iota^2}  \m + \m  \mbox{cycles}                              \m , 
\ee 
\be 
\mbox{Det}(S)      \es            (\nu_1, \nu_2, \nu_3)^2 
              \m \propto \m   \frac{{\cal V}^2}{\iota^3}                                                       \m .
\ee 
\n{\bf Remark 1} In general, such matrix elements are not geometrically independent.  
One can see them as supplying an associated linear space of conserved quantities. 
They are not moreover the only possibilities. 
\be 
A^{ij}  =  (\urho^i \cr \urho^j) 
\ee 
has equiareal \cite{Coxeter} significance. 
Being antisymmetric, this has 
\be 
\frac{n(n - 1)}{2}
\ee 
components. 
For $N = 3$, it has only one component, whereas for $N = 4$, it has 3. 
In all cases, 
\be 
\mbox{Tr}(A) = 0 
\ee 
by $A^{ij}$'s antisymmetry.  
For $N = 3$, the only other invariant is 
\be 
\mbox{det}(A)   \es   |\urho_1 \cr \urho_2|^2 
                 =    \alpha^2                  \m ,
\ee 
whereas for $N = 4$, there are two: 
\be 
\mbox{det}(A)  \es  \mbox{\Huge |}  
\s{    \mbox{\scriptsize 0}  }                            {    \s{\mbox{\scriptsize $(-\u{\rho}_1 \cr \u{\rho}_2)_{\perp}$}  }{ \mbox{(\scriptsize $-\u{\rho}_1 \cr \u{\rho}_3)_{\perp}$}  }    }  \m  
\s{    \mbox{\scriptsize $(\u{\rho}_1 \cr \u{\rho}_2)_{\perp}$}    }{    \s{\mbox{\scriptsize   0}  }                                 { \mbox{\scriptsize $(-\u{\rho}_2 \cr \u{\rho}_3)_{\perp}$}  }   }  \m
\s{    \mbox{\scriptsize $(\u{\rho}_1 \cr \u{\rho}_3)_{\perp}$}     }{    \s{  \mbox{\scriptsize $(\u{\rho}_2 \cr \u{\rho}_3)_{\perp}$}  }{ \mbox{\scriptsize 0 }   }  }     \mbox{\Huge |}         \m , 
\ee 
\be 
\mbox{II}(A)  \es  |\u{\rho}_1 \cr \u{\rho}_2|^2  \m + \m  |\u{\rho}_2 \cr \u{\rho}_3|^2  \m + \m |\u{\rho}_3 \cr \u{\rho}_1|^2   \es  \sum_{k > l} \u{\rho}^k \cr \u{\rho}^l       \m .  
\ee 
\n The presentation  
\be 
U^{ij} = E^{ij} + i \m A^{ij}
\ee
gives moreover an $n^2$ of quantities. 
These can be repackaged as one scale invariant, $\rho^2$, and an 
\be 
n^2 - 1
\ee 
of shape quantities.
In 2-$d$, the latter pick out an adjoint representation of the shape space isometry group \c{MacFarlane}, 
\be 
Isom(\mathbb{CP}^{n - 1})  \es   \frac{SU(n)}{C_n}  \m . 
\ee
This $n^2 - 1$, due to its corresponding to similarity geometry, is made by normalizing the $A_{ij}$ by $I$, so these are $\nu_i \cr \nu_j$.  
These are 3 anisoscelesnesses, 3 areas, 1 ellipticity, and a linear combination of 2 ellipicities (corresponding to the `hypercharge' \cite{Weinberg2} in the Particle Physics counterpart).

\m 

\n In 2-$d$, for $N \geq 4$ $\mI\mI(S)$ (summing over all subsystem areas) is available to form a sextet of shape quantities \cite{LR95}, 
this last quantity being one which commutes with all the other shape quantities.  

\m 

\n $N = 4$'s $\mI\mI(S)$ can also be viewed as the Casimir \cite{Gilmore} corresponding to the $SO(3)$ repackaging of the $A_{ij}$.  
This concludes our Representation Theoretic insight into volume's replacement, noting moreover that it is {\sl not} area of the whole figure but rather the 
(square root of) the sum of squares of subsystem areas.  
Thus the 'remarkable extra commuting quantity' description of this in the Molecular Physics literature \cite{LR95} 
has been clarified by a combination of basic Geometry and elementary Representation Theory.   

\m 

\n{\bf Problem 1} Observe that total area for $N \geq 4$ is not among the shape quantities.  

\m 

\n This undermines Coolidge having Shape-Theoretic significance.  
It furthermore also suggests extremizing $\mbox{II}(S)$ rather than $\alpha/\iota^2$ in generalizing \cite{I, II, III}'s Calculus considerations.  

\m 

\n For use below and in the Conclusion, the $N$-a-gon extension: $(2, \, N)$ has an 
\be 
(N - 1)^2 - 1  \es  N^2 - 2 \, N + 1 - 1  
               \es  N(N - 2)
\ee 
of $SU(N - 1)$ shape quantities.   
\be
\frac{(N - 1)(N - 2)}{2} 
\ee 
of these form an affinely significant $SO(n)$ restricted representation which carries 
\be
\frac{(N - 2)}{2} 
\ee 
further Casimirs for $N$ even, or 
\be 
\frac{(N - 1)}{2}
\ee 
for $N$ odd.  

\m 

\n General-$N$ $\lambda$-matrices, now with 
\be 
C(N - 1, \, 2)  \es  \frac{(N - 1)(N - 2)}{2}
\ee 
symmetric nondiagonal anisoscelesnesses partnering antisymetric areas, represent these.  
Let us furthermore check how many pure and how many linear combination ellipticity objects these have in general, i.e.\ what the generalization of the $J_3$ and the hypercharge are.
For $N$ even, we have 
\be 
\frac{N}{2}
\ee 
ellipticities and 
\be 
\frac{N - 2}{2}
\ee 
linear combinations of ellipticities.
On the other hand, for $N$ odd, we have 
\be 
\frac{N - 1}{2}
\ee 
ellipticities and an equal number of linear combinations of ellipticities.  

\m 

\n Problem 2) The triangle's 3 aligned 3-$d$ spaces are 6, 5 and 8-$d$ for quadrilateralland and so do not match up. 
\footnote{The Veronese embedding provides a further relation between shape quantities \cite{Kuiper, Quad-I}; 
this projective-geometric approach moreover generalizes to the Veronese--Whitney embedding for higher $N$-a-gons, 
a technique already well-known in the Shape Statistics literature  \cite{Bhatta, DM16, PE16}. 
Thus on this occasion, projective methods are more powerful than area formulae; for triangles, an area formula suffices but for quadrilaterals a projective method is needed.}

\m 

\n More generally, for the $N$-a-gon, these three numbers are
\be 
\mbox{dim}(\FrR(2, N)) = 2 \, N - 3 \m , 
\ee 
\be 
\# \, \mbox{separations}   \es  \frac{N(N - 1)}{2}
\ee 
\be 
\# \, \mbox{shape quantities} = \mbox{dim}(Isom(\mathbb{CP}^{N - 2})) = \mbox{dim}(SU(n)) = n^2 - 1 = (N - 1)^2 - 1 = N(N - 2) \m .  
\ee 
These can only coincide pairwise if one of the following holds. 
\be 
2 \, N - 3  \es  \frac{N(N - 1)}{2}  \m \Rightarrow \m  4 \, N - 6 = N^2 - N  \m \Rightarrow \m  0 = N^2 - 5 \, N + 6 = (N - 3)(N + 2) \m \Rightarrow \m N = 3            \m . 
\ee 
\be 
2 \, N - 3   =   N^2 - 2 \, N       \m \Rightarrow \m  0 = N^2 - 4 \, N + 3  =  (N - 1)(N - 3)                     \m \Rightarrow \m N = 1 \m  \mbox{ or } 3             \m . 
\ee 
\be 
\frac{N(N - 1)}{2}  \es  N(N - 2)    \m \Rightarrow \m  N^2 - N = 2 \, N^2 - 4 \, N \m \Rightarrow \m 0 = N^2 - 3 \, N = N(N - 3) \m \Rightarrow \m  N = 0 \mbox{ or } 3  \m .  
\ee 
So even the only possible pairwise coincidences have to take the triangle value $N = 3$, with occasional occurrence of the relationally trivial cases $N = 0, 1$. 
Finally, it is clear that the only full solution admitted in $N = 3$.  
In this way, the triangle case's Heron working is already unique at the basic combinatorial level.

\section{Tetrahaedron set-up and Della Francesca--Tartaglia volume formula}
%
{            \begin{figure}[!ht]
\centering
\includegraphics[width=0.35\textwidth]{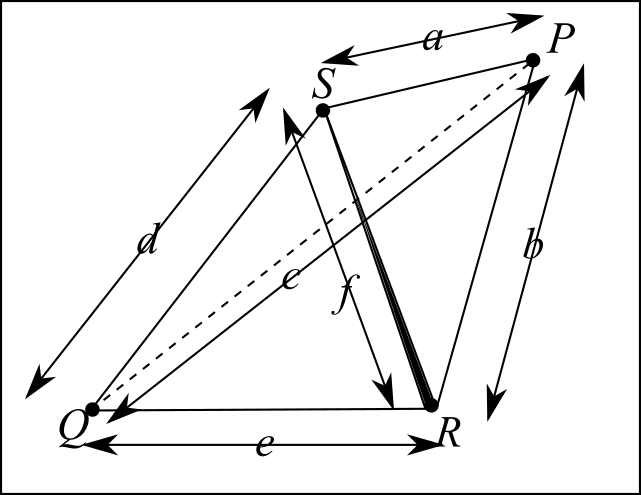}
\caption[Text der im Bilderverzeichnis auftaucht]{        \footnotesize{Tetrahaedron notation.  }  } 
\label{Tetrahaedron-Notation} \end{figure}           }

\n{\bf Definition 1} Consider an arbitrary tetrahaedron denoted as in Fig \ref{Quadrilateral-Notation}.   
We set this up such that $a$, $b$, $c$ meet at one vertex $P$ and $d$, $e$, $f$ concur pairwise at the other three vertices $Q$, $R$, $S$.  

\m 

\n We consider in particular formulae for the area $\mA$ of quadrilaterals. 

\m 

\n Tetra-area is not just the Heron quantity but also the first nontrivial Cayley--Menger quantity.
Its generalization is 
\be 
2^{d/2} d \, !\mbox{-}(d\mbox{-Volume}) =: \alpha \mbox{-}(d\mbox{-Volume}) \m .
\ee  
For $d$ even, the numerical factor $\alpha \in \mathbb{N}$, whereas for $d$ odd $\alpha \in \sqrt{2} \, \mathbb{N}$.  
\n For the tetrahaedron case, let us define 
\be 
\mU  :=  12 \sqrt{2} \, \mV          \m .   
\ee
for $\mV$ the volume.\footnote{The Introduction uses the convention that mass-weighted versions of objects are denoted in Greek letters, so ${\cal V}$ there is the mass-weighted 
analogue of $\sV$ here.}
Della Francesca--Tartaglia's formula is then as per Fig \ref{Tartaglia}.a).
Expanding out, this can moreover be rewritten as
\be 
\mU^2  =  T_{ijk} \sigma^i \sigma^j \sigma^k        \m ,
\ee 
for 
\be 
\sigma^i := (a^2, \, b^2, \, c^2, \, d^2, \, e^2, \, f^2) =: (A, \, B, \, C, \, D, \, E, \, F)
\ee
the separations$^2$ 6-vector and $T_{ijk}$ the totally-symmetric {\sl Tartaglia 3-tensor}, with components as per Fig \ref{Tartaglia}.b). 

\m 

\n Straightforwardly, there are no isotropic tensors of rank 3 in dimension 6, so questions of diagonalization are moot.
%
{\begin{figure}[ht]
\centering
\includegraphics[width=0.85\textwidth]{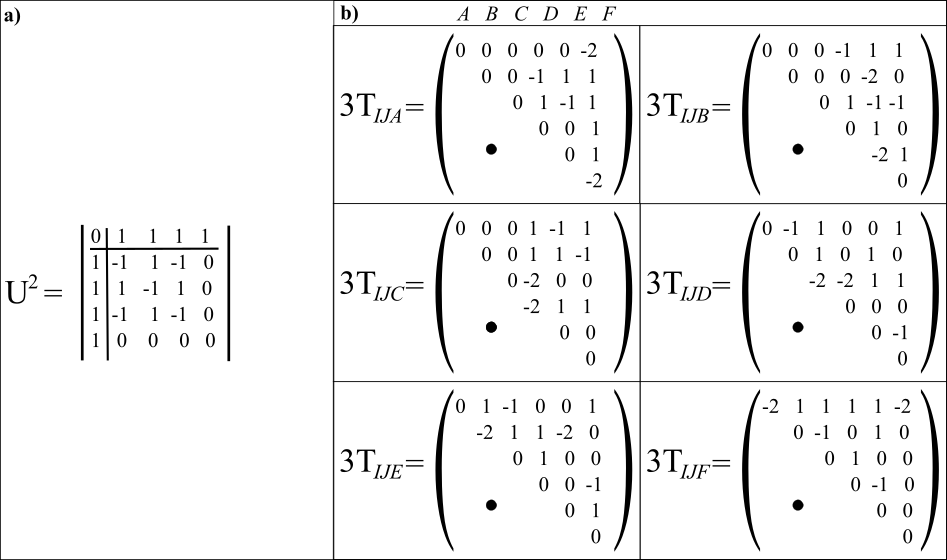}
\caption[Text der im Bilderverzeichnis auftaucht]{\footnotesize{a) Della Francesca--Tartaglia formula. 
                                                                b) the corresponding `Tartaglia 3-tensor'; 
		the heavy dot denotes that each lower triangular component coincides with the corresponding explicitly provided upper triangular component by symmetry.}} 
\label{Tartaglia}\end{figure} } 

\section{$d$-simplices and their Cayley--Menger formulae}  

\n We finally extend consideration to a $d$ simplex: we show that Heron to Cayley--Menger generalization does not extend the Little Kendall's Theorem from Heron's formula proof.  
Secs 2 and 3 cover the first two nontrivial such: $d = 2$ and 3.

\m 

\n The $d$-dimensional {\it Cayley--Menger formula} is as per Fig \ref{Cayley-Menger}.
%
{\begin{figure}[ht]
\centering
\includegraphics[width=0.35\textwidth]{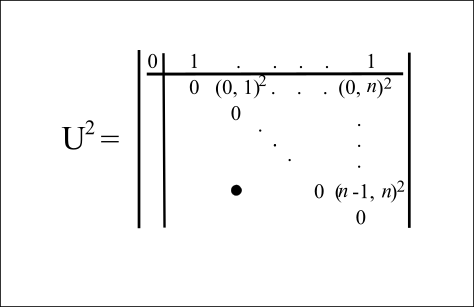}
\caption[Text der im Bilderverzeichnis auftaucht]{\footnotesize{Cayley--Menger determinant, where $(a, \, b)$ stands for distance between \cite{Blumenthal53, CH88} points $a$ and $b$.  }} 
\label{Cayley-Menger}\end{figure} } 

\m 

\n{\bf Remark 1} This can be rewritten as 
\be 
{\mU_{d}}^2 = C_{i_1 ... i_d} \sigma^{i_1} ... \sigma^{i_d} 
\ee
for  
\be 
\sigma^i = (a_1^2, ... , a_k^2)
\ee
the separations$^2$ $k$-vector, for 
\be 
k   :=          C(N, 2)  
    \es  \frac{N(N - 1)}{2}  
    \es  \frac{d(d + 1)}{2}                                            \m , 
\ee
the number of separations for $N = d + 1$ point-or-particles, and  
\be 
C_{i_1 ... i_d} 
\ee
the {\sl Cayley--Menger} $d$-array, of which (\ref{H-B}) and Fig 5.b) provide the explicit form for the first two.  

\m 

\n By appearing contracted with $d$ equal $\sigma^{i_a}$ factors, we can take the Cayley--Menger $d$-array to be totally symmetric: 
\be 
C_{i_1 ... i_d} = C_{[i_1 ... i_d]} \m . 
\ee
In forming an infinite series of totally symmetric tensors, these bear some analogy to multipole expansion tensors \cite{Jackson}. 
Those are however spatial tensors whereas Cayley--Menger tensors are configuration space tensors. 
It follows that the monopole expansion resides within a fixed dimension (usually $d = 3$), whereas Cayley--Menger tensors increase in dimension according to 
\be 
\frac{N(N - 1)}{2} \es \frac{d(d + 1)}{2}   \m .
\ee 
\n We are to next contemplate diagonalizing these, i.e.\ solving 
\be 
C_{i_1 ... i_d}  =  \lambda \, I_{i_1 ... i_d}                         \m .  
\label{gen-eval}
\ee 
for $I_{i_1 ... i_d}$ the $d$-dimensional matchingly totally symmetric isotropic tensor. 
This gives separate even and odd cases.

\m 

\n On the one hand, for odd $d$, there are no rank-$d$ isotropic tensors in dimension $k$, so (\ref{gen-eval}) is moot.  

\m

\n On the other hand, for even-$d$ (> 0), we do have an eigenvalue problem, producing 
\be 
\left(\frac{d(d + 1)}{2}\right)^{d/2}
\ee 
eigenvalues.
This is to be compared with the Casson sphere corresponding dimensionally to 
\be 
\frac{d(d + 1)}{2}
\ee 
quantities summing to 1. 
[$\iota$ remains available as a scale quantity along the Casson diagonal.]
So, for $d > 0$ even, we would require 
\be 
\left(\frac{d(d + 1)}{2}\right)^{d/2}  \es  \frac{d(d + 1)}{2} \m . 
\ee  
I.e.\ 
\be
\left(\frac{d(d + 1)}{2}\right)^{d/2 - 1} = 1
\ee 
which can only be solved for $d \in 2\mathbb{N}$ by 
\be 
d = 2                                                      \m .
\ee 
Thus, even just combinatorially, the extent of the Cayley--Menger--Casson coincidence is just the Heron--Kendall--Casson coincidence yielding the 2-sphere of triangles.

\section{Conclusion}

It was recently shown \c{A-Heron} that for triangle constellations in 2-$d$, the form taken by the corresponding shape space -- a sphere -- can be derived from Heron's formula. 
This is a fourth derivation of this result: `Kendall's Little Theorem'.  
The other three are Kendall's extremization \c{Kendall} and using the Hopf map (both outlined in the Appendix), 
and setting up an indirectly-formulated similarity mechanics action for the problem and conducting Lagrangian-level reduction \c{FORD} (or similar \c{FileR, ABook}).
The Hopf map itself is moreover derived from Heron's formula in the process. 
Its three coordinate functions -- mass-weighted tetra-area, anisoscelesness and ellipticity 
in the shape theoretic context -- arise as the subject of Heron's formula and the two non-unit eigenvectors of the Heron--Buchholz matrix.

\m 

\n In the current article, however, we show that this fourth derivation is a one-off, 
firstly in the sense that not even quadrilateral constellations admit an analogous derivation from an area formula. 
While the passage from Brahmagupta's formula to Bretschneider's and then to Coolidge's is shown to be progressively more suitable for Shape Theory, 
even Coolidge's formula fails to yield a comparable result.
One underlying reason for this is that the shape space of quadrilaterals is $\mathbb{CP}^2$ (and that of $N$-a-gons is $\mathbb{CP}^{N - 2}$ more generally: Kendall's Theorem). 
This is somewhat more structurally complex than $\mathbb{CP}^1 = \mathbb{S}^2$, 
so some methods of arriving at this 2-$d$ space are `spherical' rather than `projective' and so fail to generalize to projective cases.
All the other three derivations mentioned above moreover do generalize to $N$-a-gons, with the Hopf case proceeding via one of the usual generalizations of the Hopf map, 
as per Fig 1.c-d).
The $N$-a-gon versions of these derivations are covered likewise in the Appendix and \c{Kendall, FORD, FileR, ABook}.  
Another underlying reason is because total area of a figure (after mass-weighting and dividing by the matching power of the moment of inertia) 
does not constitute a shape quantity in 2-$d$ for $N \geq 4$. 
It is, rather, the sum of squares of the triangle subsystems supported by a given choice of Jacobi coordinates which, upon being dressed in this manner, 
constitutes a shape quantity \c{LR95}.
In the process of analyzing this, we render clear that this  'remarkable extra commuting quantity' -- as described in the Molecular Physics literature \cite{LR95} --
has been given a clear explanation rooted in both basic Geometry and elementary Representation Theory.   

\m 

\n Some dimensional coincidences supporting special features of the triangle case, and its Heron derivation, were also provided in Sec 4. 
These rely on scaled triangles having 3 independent relational coordinates, 3 relative separations and 3 shape quantities. 
Moreover, the relational space of scaled triangles is (conformal to) $\mathbb{R}^3$: the space in which spheres are most commonly modelled as extrinsically residing within. 
But these threes become 
\be 
2 \, N - 3 \mma \frac{N(N - 1)}{2} \mma N(N - 2) 
\ee 
for $N$-agons, for which we showed that even just double coincidences imply that $N = 3$, or the relationally trivial 0 or 1, and $N = 3$ is the sole triple solution.  

\m 

\n We also explained that the triangeland $\mathbb{S}^2$ is also the first nontrivial simplexland alias basisland: the 
\be 
N = d + 1
\ee 
diagonal which Casson showed to be topologically $\mathbb{S}^{d(d + 1)/2 - 1}$.  
Thus `Kendall's Little Theorem' may additionally be viewed as the unique case in which both Kendall's Theorem and Casson's Theorem apply. 
This means that whether derivations of  `Kendall's Little Theorem' admit Casson's Theorem generalizations should also be checked.  

\m 

\n The next case along on the simplexland diagonal is tetrahaedronland. 
This admits the della Francesca--Tartaglia volume formula, 
as is most easily seen from recasting Heron's formula as a determinant and generalizing dimensionally.  
This can be reformulated in terms of a (new, as far as the Author is aware) `totally symmetric Tartaglia 3-tensor' 
in the 6-$d$ separation space supported by tetrahaedral constellations.
This however does not admit eigenvalues and eigenvectors since there are no isotropic tensors of rank 3 in 6-$d$.  

\m 

\n Our next observations are that the entire simplexland diagonal is populated by the hypervolume formulae of Cayley--Menger \cite{Cayley, Menger, Blumenthal53, CH88}. 
That these can be reformulated in terms of for the (also new, as far as the Author is aware) infinite series of Cayley--Menger totally-symmetric $d$-tensors. 
That mass-weighted $d$-hypervolume divided by a matching power of the moment of inertia is a suitable shape quantity for each $d$. 
That the even-$d$ cases among these do possess the right dimensions-and-rank isotropic tensors to have eigenvalue problems. 
These eigenvalue problems, however, do not have the right dimensionality to produce on-$\mathbb{S}^{d(d + 1)/2 - 1}$ conditions, other than in the $d = 2$ case 
in which the Heron derivation of `Kendall's Little Theorem' is recovered.  

\m 

\n \c{A-Heron} and the current article demonstrate that combining basic undergraduate Geometry and Linear Algebra is still capable of producing new results.  
Taking the stance that Shape Theory is a futuristic branch of both Statistics and Background-Independent Physics, 
we note moreover that maximally straightforward derivations of its simplest results is of considerable {\sl pedagogical} significance.  
Heron's formula is widely known in schools across the globe, and at least in the U.K., 
eigenvalues and eignevectors are within what is taught to mathematically-inclined high school students in their final year.
So a derivation of `Kendall's Little Theorem' that the space of triangles is a sphere based on just these elements 
renders Shape Theory presentable to mathematically-inclined high-school students.  
Kendall's extremization in turn lies within the grasp of undergraduates majoring in mathematics, 
whereas the Hopf map and Mechanics reduction workings are more commensurate with what is covered in graduate school 
(be that Geometry, Geometrical Methods, or a partly geometrical course in Dynamics or Statistics).  
It is therefore interesting to see how far one can take each route to Shape Theory, 
with the current Article revealing that the technically-simplest of all known derivations 
of a nontrivially shape-theoretic result {\sl not} extending beyond the first case of triangles. 
The current article thus represents a {\sl caveat} in the most introductory exposition of Shape Theory, that while knowing Heron's formula 
and eigenvalue--eigenvector workings suffices to obtain the triangleland sphere, other kinds of derivations are required for the larger shape spaces.    

\m 

\n Let us finally note that the Hopf map generalizing along the $N$-a-gonlands plays a deep underpinning role in these being more geometrically understood 
than the simplexland diagonal. 
This is with particular reference to Kendall's Theorem being a {\sl metric} as well as topological result, whereas Casson's Theorem is just topological.  
This, and accumulated knowledge about the metric geometry and associated linear methods for $\mathbb{CP}^{N - 2}$, makes $N$-a-gons far more amenable 
to physical study than simplexlands. 
There is one sense in which this is unfortunate: that tetrahadreons -- the 4-body problem in 3-$d$ -- is more directly relevant to nature than $N$-a-gons in 2-$d$. 
One sense in which this does not matter: $N$-a-gons will do as a model arena of many aspects of classical and quantum GR's dynamics, problem of time 
and background independence.
And one sense in which this is fortunate: that $N$-a-gons provide a `shape representation' for quantum-information-theoretic qu-$N$its \c{Brody, CGP15}.   

\m 

\n{\bf Acknowledgments} I thank Chris Isham and Don Page for previous discussions.  
Reza Tavakol, Malcolm MacCallum, Enrique Alvarez and Jeremy Butterfield for support with my career.  

\begin{appendices}

\section{Kendall extremization and Hopf fibration projection proofs}

Preshape space is straightforwardly a sphere (\ref{Preshape}). 
This carries moreover the extrinsically defined chordal metric or the topologically equivalent \cite{Kendall} intrinsically defined great circle metric, 
\be 
D(A, \, B) \es  \mbox{arccos}(A, \, B)
\ee 
2-$d$ shape space $\FrS(2, \, N)$ then carries the quotient metric 
\be 
D(Q(A, \, B)  \es \s{min}{R \, \in \, SO(2)} D( A, \, R \, B) \es \s{min}{R \, \in \, SO(2)} \mbox{arccos}( A, \, R \, B) \m .  
\ee 
Kendall then shows by carrying out the corresponding basic Calculus extremization that 
\be 
\mbox{cos} \, D(Q(\biz), Q(\biw)) \es \frac{|(\biz \cdot \biw)_{\sC}|}{||\biw||_{\sC} ||\biz||_{\sC}}  
\label{Lemma}
\ee 
for 
\be 
(\biw \cdot \biz )_{\sC} := \sum_{i = 1}^n z_A \bar{w}^A
\ee 
and $|| \mbox{ } ||_{\sC}$ the corresponding norm.
Then a small perturbation 
\be 
\biw = \biz + \delta \biw
\ee 
brings about a small change 
\be 
\delta D^2 = \mbox{sin}^2 \delta D + O((\delta D)^4) = 1 - \mbox{cos}^2 D(Q(\biz), \, Q(\biz + \delta \biz) + O((\delta\biz)^4)
\ee 
so using (\ref{Lemma})
\be 
= 1 - \frac{|(\biz \cdot (\biz + \delta \biz)_{\sC}|^2}{||\biz||^2_{\sC} ||\biz + \delta \biz||^2_{\sC}}  
\ee
so expanding 
\be 
= \frac{||\biz||_{\sC}||\delta \biz||_{\sC}  - |(\biz \cdot \delta \biz)_C|^2}{||\biz||^4_{\sC}}  +  O((\delta\biz)^4) \m . 
\ee  
Finally take the limit as $\delta \biz \longrightarrow 0$ to obtain the  Fubini--Study metric in standard homogeneous coordinates, 
\be 
\d s^2 \es = \frac{||\biz||_{\sC}||\d \biz||_{\sC}  - |(\biz \cdot \d \biz)_C|^2}{||\biz||^4_{\sC}}   \m .  
\ee 
Changing homogeneous coordinate patch when necessary, it can be ascertained that this recovers the entirety of $\mathbb{CP}^{N - 2}$.

\mbox{ } 

\n If one proceeds the Hopf way instead, the Fubini--Study metric arises by projection along the fibres: i.e. now a bundle-theoretic construct.\footnote{See e.g.\ 
\c{Nakahara, CGP15} for obtaining the Fubini--Study metric from K\"{a}hler first principles. 
There is moreover no known shape theoretic first principle for the N-a-gonland's K\"{a}hler potential.}
%
In either case, for a single $z$ the Fubini--Study metric collapses to 
\be 
\d s^2 = \frac{|\d z|^2}{(1 + |z|^2)^2}                  \m , 
\ee 
Then using the polar coordinates representation 
\be 
z = r \, \mbox{e}^{i \phi}                               \m , 
\ee 
we obtain 
\be 
\d s^2 \es  \frac{\d r^2 + r^2 \d\phi^2}{4(1 + r^2)^2}   \m , 
\ee 
which is readily identified as the sphere in stereographic polar coordinates. 
Finally using the venerable substitution
\be
r \es \mbox{tan} \, \mbox{$\frac{\theta}{2}$}  \m , 
\ee
the standard spherical coordinates form of the metric, 
\be 
\d s^2 = \d \theta^2 + \mbox{sin}^2\theta \, \d \phi^2    \m , 
\ee 
is recovered up to constant proportion. 

\end{appendices}



\begin{thebibliography}{99}

\footnotesize


\bibitem{Hero}                Heron, alias Hero, of Alexandria, {\it Metrica} (60 A.D.)

\bibitem{Ptolemy}             C. Ptolemy, {\it Almagest} Book 1 (2nd Century A.D.); 
                              for an English translation, see e.g.\ G.J. Toomer, {\it Ptolemy's Almagest} 2nd ed. (1998).

\bibitem{Brahmagupta}         Brahmagupta, (7th Century A.D.)

\bibitem{DF}                  P. della Francesca (15th Century A.D.)

\bibitem{Tartaglia}           N. Fontana alias Tartaglia (16th Century A.D.)
						  
\bibitem{Cayley}              A. Cayley, "A Theorem in the Geometry of Position," Cambridge Math.", {\bf 2}, 267 (1841); 
                              This is more accessibly published in {\it Collected works of Arthur Cayley} (Cambridge University Press, Cambridge 2009).

\bibitem{Bretschneider}       C.A. Bretschneider, ``Untersuchung der Trigonometrischen Relationen des Geradlinigen Viereckes" 
                              (Investigation of the Trigonometric Relations of Quadrilaterals) Archiv. der Math. {\bf 2} 225 (1842).							  
							  							  
\bibitem{Veronese}            See e.g. J. Harris, {\it Algebraic Geometry. A First Course} (Springer-Verlag, New York 1992); 

                              G. Veronese originally wrote on such matters in the 1880's.  
							  

\bibitem{Menger}              K. Menger, ``Untersuchungen Ueber Allgemeine Metrik", Math. Ann. {\bf 100} 75 (1928); 

                                         ``New Foundation of Euclidean Geometry," Am. Math. {\bf 53} 721 (1931). 
										   
\bibitem{Hopf}                H. Hopf, ``\"{U}ber die Abbildungen der dreidimensionalen Sph\"{a}re auf die Kugelfl\"{a}che", 
                              (``Concerning the Images of $\mathbb{S}^3$ on $\mathbb{S}^2$"), Math. Ann. (Berlin) Springer {\bf 104} 637 (1931).  
						
\bibitem{Coolidge}            J.L. Coolidge, ``A Historically Interesting Formula for the Area of a Quadrilateral", Amer. Math. Monthly {\bf 46} 345 (1939).

\bibitem{Blumenthal53}        L. Blumenthal, {\it Theory and Applications of Distance Geometry} (Cambridge University Press, Cambridge 1953; reprinted by Chelsea, New York 1970). 

\bibitem{CG67}                H.S.M. Coxeter and S.L. Greitzer, {\it Geometry Revisited} (Mathematical Association of America, 167).  

\bibitem{DeWitt}			  B.S. DeWitt, ``Quantum Theory of Gravity. I. The Canonical Theory.", Phys. Rev. {\bf 160} 1113 (1967). 
 					
\bibitem{Battelle}            J.A. Wheeler, in {\it Battelle Rencontres: 1967 Lectures in Mathematics and Physics} ed. C. DeWitt and J.A. Wheeler (Benjamin, New York 1968).  
	
\bibitem{Kuiper}              N.H. Kuiper, ``The Quotient Space of $\mathbb{CP}^2$ by Complex Conjugation is the 4-Sphere" Math. Ann. {\bf 208} 175 (1974).  
	
						
\bibitem{Kendall84}           D.G. Kendall, ``Shape Manifolds, Procrustean Metrics and Complex Projective Spaces", Bull. Lond. Math. Soc. {\bf 16} 81 (1984). 

\bibitem{CH88}                G. Crippen and T.F. Havel, {\it Distance Geometry and Molecular Conformation} (1988).  

\bibitem{Coxeter}             H.S.M. Coxeter, {\it Introduction to Geometry} (Wiley, New York 1989).

\bibitem{Kendall89}           D.G. Kendall, ``A Survey of the Statistical Theory of Shape", Statistical Science {\bf 4} 87 (1989).

							  
\bibitem{Marchal}             C. Marchal, {\it Celestial Mechanics} (Elsevier, Tokyo 1990). 

\bibitem{Nakahara}            M. Nakahara, {\it Geometry, Topology and Physics} (Institute of Physics Publishing, London 1990).   

\bibitem{Buchholz}            R.H. Buchholz, ``Perfect Pyramids", Bull. Austral. Math. Soc. {\bf 45} 353 (1992). 

\bibitem{Kuchar92}            K.V. Kucha\v{r}, ``Time and Interpretations of Quantum Gravity", 
                              in {\it Proceedings of the 4th Canadian Conference on General Relativity and Relativistic Astrophysics} 
                              ed. G. Kunstatter, D. Vincent and J. Williams (World Scientific, Singapore, 1992),
%
                              reprinted as Int. J. Mod. Phys. Proc. Suppl. {\bf D20} 3 (2011). 
					  
\bibitem{I93}                 C.J. Isham, ``Canonical Quantum Gravity and the Problem of Time", 
                              in {\it Integrable Systems, Quantum Groups and Quantum Field Theories} 
                              ed. L.A. Ibort and M.A. Rodr\'{\i}guez (Kluwer, Dordrecht 1993), gr-qc/9210011.

\bibitem{LR95}                R.G. Littlejohn and M. Reinsch, ``Internal or Shape Coordinates in the $N$-Body Problem", Phys. Rev. {\bf A52} 2035 (1995); 

\bibitem{Weinberg2}           S. Weinberg, {\it The Quantum Theory of Fields. Vol II.  Modern Applications.}  (Cambridge University Press, Cambridge 1995).  

\bibitem{LR97}                R.G. Littlejohn and M. Reinsch,  ``Gauge Fields in the Separation of Rotations and Internal Motions in the $N$-Body Problem", 
                              Rev. Mod. Phys. {\bf 69} 213 (1997); 
							  
							  K.A Mitchell and R.G. Littlejohn, ``Kinematic Orbits and the Structure of the Internal Space for Systems of Five or More Bodies",
                              J. Phys. A: Math. Gen. {\bf 33} 1395 (2000).  

\bibitem{Small}               C.G.S. Small, {\it The Statistical Theory of Shape} (Springer, New York, 1996).  

\bibitem{Jackson}             J.D. Jackson, {\it Classical Electrodynamics}, (Wiley, Chichester 1998).  

\bibitem{Kendall}             D.G. Kendall, D. Barden, T.K. Carne and H. Le, {\it Shape and Shape Theory} (Wiley, Chichester 1999).  


\bibitem{Cohn}                P.M. Cohn, {\it Classic Algebra} (Wiley, Chichester 2000).  

\bibitem{JM00}                K.V. Mardia and P.E. Jupp, {\it Directional Statistics} (Wiley, Chichester 2000).

\bibitem{Montgomery}          R. Montgomery, ``Infinitely Many Syzygies", Arch. Rat. Mech. Anal. {\bf 164} 311 (2002); 

                                              ``Fitting Hyperbolic Pants to a 3-Body Problem", Ergod. Th. Dynam. Sys. {\bf 25} 921 (2005), math/0405014; 
											  
											  ``The Three-Body Problem and the Shape Sphere",  Amer. Math. Monthly {\bf 122} 299 (2015), arXiv:1402.0841.   							  
		
\bibitem{MacFarlane}          A.J. MacFarlane, ``Complete Solution of the Schr\"{o}dinger Equation of the Complex Manifold $\mathbf{CP}^2$", 
                              J. Phys. A: Math. Gen. {\bf 36} 7049 (2003); 

                              ``Solution of the Schr\"{o}dinger equation of the complex manifold $\mathbb{CP}^n$", 
                              J. Phys. A: Math. Gen. {\bf 36} 9689 (2003).  
		
\bibitem{Brody}               D.C. Brody, Shapes of Quantum States, J. Phys. {\bf A37} 251 (2004), quant-ph/0306013.
				
\bibitem{Gilmore}             R. Gilmore, {\it Lie Groups, Lie Algebras, and Some of Their Applications} (Dover, New York 2006).  
  			
\bibitem{FORD}                E. Anderson, ``Foundations of Relational Particle Dynamics", Class. Quant. Grav. {\bf 25} 025003 (2008), arXiv:0706.3934; 

                                           ``Six New Mechanics corresponding to further Shape Theories", Int. J. Mod. Phys. {\bf D 25} 1650044 (2016), arXiv:1505.00488. 
          	 
\bibitem{GT09}                D. Groisser, and H.D. Tagare, ``On the Topology and Geometry of Spaces of Affine Shapes", 
                              Journal of Mathematical Imaging and Vision {\bf 34} 222 (2009).  
							  		
\bibitem{+Tri}                E. Anderson, ``Shape Space Methods for Quantum Cosmological Triangleland", Gen. Rel. Grav. {\bf 43} 1529 (2011), arXiv:0909.2439.  

			
\bibitem{Quad-I}              E. Anderson, ``Relational Quadrilateralland. I. The Classical Theory", Int. J. Mod. Phys. {\bf D23} 1450014 (2014), arXiv:1202.4186; 

                              E. Anderson and S.A.R. Kneller, ``Relational Quadrilateralland. II. The Quantum Theory", Int. J. Mod. Phys. {\bf D23} 1450052 (2014), arXiv:1303.5645.  
				
\bibitem{APoT}                E. Anderson, ``The Problem of Time in Quantum Gravity", in {\it Classical and Quantum Gravity: Theory, Analysis and Applications}  
                              ed. V.R. Frignanni (Nova, New York 2012), arXiv:1009.2157; 

							               ``Problem of Time in Quantum Gravity", Annalen der Physik, {\bf 524} 757 (2012),  arXiv:1206.2403;     

                                           ``Beables/Observables in Classical and Quantum Gravity", SIGMA {\bf 10} 092 (2014), arXiv:1312.6073; 

                                           ``Explicit Partial and Functional Differential Equations for Beables or Observables" arXiv:1505.03551; 

                                           ``On Types of Observables in Constrained Theories", arXiv:1604.05415.  
				   												
\bibitem{FileR}               E. Anderson, ``The Problem of Time and Quantum Cosmology in the Relational Particle Mechanics Arena", arXiv:1111.1472.  

\bibitem{Bhatta}              A. Bhattacharya and R. Bhattacharya, {\it Nonparametric Statistics on Manifolds with Applications to Shape Spaces} 
                             (Cambridge University Press, Cambridge 2012).

\bibitem{APoT3}               E. Anderson, ``Problem of Time and Background Independence: the Individual Facets", arXiv:1409.4117.  
							 							 
\bibitem{MIT}                 A. Edelman and G. Strang, ``Random Triangle Theory with Geometry and Applications", 
                              Foundations of Computational Mathematics (2015), arXiv:1501.03053.   

\bibitem{CGP15}               M. Cvetic, G.W. Gibbons and C.N. Pope, ``Compactifications of Deformed Conifolds, Branes and the Geometry of Qubits", 
                              JHEP {\bf 135} 01(2016), arXiv:1507.07585.
							  
\bibitem{PE16}                V. Patrangenaru and L. Ellingson ``Nonparametric Statistics on Manifolds and their Applications to Object Data Analysis" 
                             (Taylor and Francis, Boca Raton, Florida 2016).  

\bibitem{KKH16}               F. Kelma, J.T. Kent and T. Hotz, ``On the Topology of Projective Shape Spaces", arXiv:1602.04330. 

\bibitem{DM16}                I.L. Dryden, K.V. Mardia, {\it Statistical Shape Analysis: With Applications in R}, 2nd Edition (Wiley, Chichester 2016).  
								 
\bibitem{ABook}               E. Anderson, {\it The Problem of Time}, Fundam.Theor.Phys. {\bf 190} (2017) pp.- ; 

                              alias {\it The Problem of Time. Quantum Mechanics versus General Relativity}, (Springer, New York 2017); 

                              its extensive Appendix Part ``Mathematical Methods for Basic and Foundational Quantum Gravity",           
							  is freely accessible at https://link.springer.com/content/pdf/bbm$\%$3A978-3-319-58848-3$\%$2F1.pdf ; 
							  
							  ``A Local Resolution of the Problem of Time", arXiv:1809.01908.  
	
\bibitem{I}                   E. Anderson, ``The Smallest Shape Spaces. I. Shape Theory Posed, with Example of 3 Points on the Line", arXiv:1711.10054.

\bibitem{II}                   E. Anderson,                                           ``The Smallest Shape Spaces. II. 
                                             4 Points on a Line Suffices for a Complex Background-Independent Theory of Inhomogeneity", arXiv:1711.10073.
					  
\bibitem{III}                 E. Anderson, ``The Smallest Shape Spaces. III. Triangles in the Plane and in 3-$d$", arXiv:1711.10115. 

\bibitem{A-Heron}             E. Anderson, ``Two New Perspectives on Heron's Formula",  arXiv:1712.01441.  

\bibitem{Shape-Theory}                      ``Alice in Triangleland: Lewis Carroll's Pillow Problem and Variants Solved on Shape Space of Triangles", arXiv:1711.11492;
 
                                           ``Maximal Angle Flow on the Shape Sphere of Triangles", arXiv:1712.07966; 

                                           ``Monopoles of Twelve Types in 3-Body Problems", arXiv:1802.03465; 
										   
										   ``Topological Shape Theory", arXiv:1803.11126;    

                                           ``Background Independence: $\mathbb{S}^1$ and $\mathbb{R}$ Absolute Spaces differ greatly in Shape-and-Scale Theory", arXiv:1804.10933; 
							   
							               ``Rubber Relationalism: Smallest Graph-Theoretically Nontrivial Leibniz Spaces", arXiv:1805.03346; 
										   
                                           ``Absolute versus Relational Debate: a Modern Global Version", arXiv:1805.09459;

                                           ``$N$-Body Problem: Minimal $N$ for Qualitative Nontrivialities", arXiv:1807.08391; 

										   ``Isotropy Groups and Kinematical Orbits for 1 and 2-$d$ $N$-Body Problems", arXiv:0810.04043;  

								           ``$N$-Body Problem: Minimal $N$ for Qualitative Nontrivialities II: Varying Carrier Space and Group Quotiented Out", forthcoming November 2018.  
										   									  
\bibitem{Tri-Ineq}            E. Anderson, ``Shape (In)dependent Inequalities for Triangleland's Jacobi and Democratic-Linear Ellipticity Quantitities", arXiv:1712.04090.  

\bibitem{VW}                  Y. Wang and V. Patrangenaru ``Nonparametric Inference for Location Parameters of Veronese Whitney means and antimeans on Kendall Shape Spaces", 
                              arXiv:1806.08683

\bibitem{PE-1}                E. Anderson, ``Specific PDEs for Preserved Quantities in Geometry. I. Similarities and Subgroups", arXiv:1809.02045.    

\bibitem{AObs4}               E. Anderson,  ``Spaces of Observables from Solving PDEs. I. Translation-Invariant Theory.", arXiv:1809.07738.   

\bibitem{AMink}               E. Anderson, ``Event Shapes: Shape Theory in Minkowski Spacetime", forthcoming October 2018. 

\bibitem{DO-2}                E. Anderson,  ``Solving PDEs for Spaces of Observables. II. Preshape Theory, forthcoming September 2018.								
			
\bibitem{IV}                  E. Anderson, ``Quadrilaterals in Shape Theory. I. Shapes, Coordinates and Shape Spaces", forthcoming November 2018.  
		
\bibitem{Quad-Sub}            E. Anderson, ``Quadrilaterals in Shape Theory. V. Geometrically Significiant Submanifolds of Quadrilaterals", forthcoming November 2018.  

\bibitem{A-Quad-Ineq}         E. Anderson, ``Quadrilaterals in Shape Theory. III. Shape-Theoretic Inequalities", forthcoming.
							  													
\end{thebibliography}
\end{document}